# Gene regulatory network inference algorithm based on spectral signed directed graph convolution

Rijie Xi, Weikang Xu, Wei Xiong, Yuannong Ye and Bin Zhao

*Abstract*—Accurately reconstructing Gene Regulatory Networks (GRNs) is crucial for understanding gene functions and disease mechanisms. Single-cell RNA sequencing (scRNA-seq) technology provides vast data for computational GRN reconstruction. Since GRNs are ideally modeled as signed directed graphs to capture activation/inhibition relationships, the most intuitive and reasonable approach is to design feature extractors based on the topological structure of GRNs to extract structural features, then combine them with biological characteristics for research. However, traditional spectral graph convolution struggles with this representation. Thus, we propose MSGRNLink, a novel framework that explicitly models GRNs as signed directed graphs and employs magnetic signed Laplacian convolution. Experiments across simulated and real datasets demonstrate that MSGRNLink outperforms all baseline models in AUROC. Parameter sensitivity analysis and ablation studies confirmed its robustness and the importance of each module. In a bladder cancer case study, MSGRNLink predicted more known edges and edge signs than benchmark models, further validating its biological relevance.

## I. INTRODUCTION

Gene regulatory networks (GRNs) consist of intricate regulatory interactions between transcription factors (TFs) and target genes, which is a crucial mechanism for maintaining life processes, controlling biochemical reactions, and regulating the levels of chemical compounds. GRNs play a key role in many domains, such as gene function prediction, cancer biomarkers identification, and the discovery of potential drug targets [1]. The development of gene sequencing technology has facilitated data acquisition and produced a wealth of gene microarray data. Particularly, single-cell RNA sequencing technology (scRNA-seq) has deepened insights into cellular heterogeneity, providing opportunities to identify high-resolution transcriptional states and transitions [2]. In recent years, researchers have devised many supervised and unsupervised computational methods to deduce GRN from scRNA-seq data, which are primarily classified into three categories: information theory-based methods, machine learning-based methods, and deep learning-based methods.

Methods based on information theory suggest that genes within the same group have analogous expression patterns during physiological processes and forecast regulatory linkages by assessing the correlations among genes [3]. For instance, Chan et al. developed the undirected unsigned model PIDC [4] in 2017 that uses partial information decomposition (PID) to uncover gene regulatory relationships. In the same year, Specht et al. presented the undirected unsigned model LEAP [5], which infers GRNs by calculating Pearson correlations on fixed size time windows with different lags. Later, in 2020, Aibar et al. proposed the undirected unsigned SCRIBE [6], a model that constructs GRNs based on the mutual information between the past state of a regulator and the current state of a target gene. These methods have the advantage of requiring a small sample size and having minimal computing cost, which allows for the construction of massive networks from little quantities of data. However, the GRNs inferred by this kind of approach are undirected and fail to distinguish between upstream and downstream of the regulatory link because the correlations used in the aforementioned literature are bidirectional [7]. Furthermore, these approaches ignore the known network topology and only employ gene expression profiles.

Machine learning-based algorithms transform the GRN inference problem into a classification or regression problem. For example, Huynh-Thu et al. introduced the undirected unsigned model GENIE3 [8] in 2010, which decomposes the prediction of GRN between *p* genes into *p* distinct regression problems. In 2017, Hirotak et al. proposed the directed unsigned model SCODE [9], which integrates linear ordinary differential equations and linear regression to infer GRNs. Later, in 2019, Moerman et al. proposed the undirected unsigned model GRNBoost2 [10], an efficient algorithm based on gradient boosting within the GENIE3 [8] framework. In 2020, Ghosh et al. employed Lasso to introduce an ensemble regression algorithm PoLoBag [11] — this was the first model in GRN inference to simultaneously consider both sign and direction. Most Recently, in 2022, Abdullah et al. designed a non-convex optimization model within the ADMM framework and proposed scSGL [12], which further incorporates kernel functions to model undirected signed gene regulatory

This work was supported by National Natural Science Foundation of China (12001465, 12261086) and by Guizhou Provincial Basic Research Program (Natural Science Category) Project ZK [2023]298. (*Corresponding author: Bin Zhao*).

Bin Zhao and Yuannong Ye are with the School of Biology and Engineering, Guizhou Medical University, Guiyang 550004, China (E-mail: scott84@sina.com; yyn@gmc.edu.cn).

Rijie Xi, Weikang Xu and Wei Xiong are with the School of Mathematics and Systems Science, Xinjiang University, Urumqi 830000, China (e-mail: rijiexi06@gmail.com; xuweikang34@gaiml.com; xingheng-1985@163.com).

Our source code for this work is available at https://github.com/XRJ0629/MSGRNLink.

Color versions of one or more of the figures in this article are available online at http://ieeexplore.ieee.org

networks.

Deep learning-based methods aim to process raw biological data to infer GRNs using classical deep learning algorithms. For example, Kishan et al. proposed the undirected unsigned model GNE [13] in 2019, a supervised model that uses MLP to encode gene expression for predicting the interactions between genes. In the same year, Yuan et al. developed CNNC [14], a directed unsigned model that transforms gene pair co-expression into image-like histograms and applies CNNs for classification. Although both GNE [13] and CNNC [14] incorporate gene expression profile information and network topology, they cannot handle time series data. To address this limitation, DGRNS [15] was created in 2022, combining RNN and CNN to capture temporal and spatial features respectively, thereby enabling accurate inference of directed unsigned regulatory relationships among genes.

Traditional deep learning methods are not well-suited for non-Euclidean data such as GRN, as they fail to handle the topology of the network. To address this limitation, an increasing number of studies have adopted Graph Neural Network (GNN) for GRN inference. In 2020, Wang et al. introduced GRGNN [16], the first model to apply GNN to GRN inference. It is an undirected unsigned model that trains GNN classifiers using positive and negative subgraphs. In 2022, Chen et al. proposed a directed unsigned model GENELink [17], which first reconstructs GRN employing GAT and predicts potential interactions through a self-attention mechanism. GENELink [17] addresses the issue raised in GNE [13], where capturing topological information using one-hot gene ID vector is cumbersome. However, GENELink [17] suffers from issues such as high computational complexity and performance degradation due to overemphasis on higher-order neighbors. To overcome these drawbacks, in 2024, GATCL [18] enhanced the use of GAT in GRN inference by integrating multi-head and single-head attention layers, thereby improving both efficiency and predictive accuracy. In 2023, Wang et al. developed DeepRIG [19], another undirected unsigned model that constructs a prior regulatory graph by calculating Spearman correlation coefficients between gene pairs and performs inference using Graph Auto-Encoder (GAE). In the same year, Mao et al. proposed GNNLink [20], a directed unsigned model that incorporates a GNN-based interactive graph encoder and infers GRN by performing matrix complementation on node features. Also in 2023, Wei et al. introduced DGCGRN [21], a directed unsigned model that utilizes a directed graph convolutional network (DGCN). In DGCGRN [21], spectral convolution is performed in its most basic form by directly multiplying the normalized adjacency matrix with the node feature matrix. To avoid the disadvantage of poor performance of traditional convolution on directed graphs, so DGCGRN [21] applied this convolution to the first-order approximate matrix, the second-order in-degree approximate matrix, and the second-order out-degree approximate matrix separately to considering direction in GRN. That is, DGCGRN [21] does not start from the most fundamental design of convolutional kernel to apply to directed graphs. Moreover, this model is unable to distinguish the regulatory sign of GRN. Later, in 2024, Gao et al. proposed DeepFGRN [22], a directed signed model for inferring GRN from bulk gene expression data. It features a bidirectional node representation module based on Generative Adversarial Network (GAN) with dual generators. That same year, Liu et al. presented CVGAE [23], a directed unsigned model that employs GNN for inductive representation learning and predicts gene interactions based on the mathematical distances between genes. In 2025, Yu et al. proposed GCLink [24], a directed unsigned model that leverages graph augmentation and contrastive learning to reduce reliance on labeled data. Meanwhile, Kommu et al. developed ScRegNet [25], which combines single-cell foundation models (scFMs) with GNN through pre-training and transfer learning to infer GRN.

TABLE I
CURRENT GRN INFERENCE ALGORITHMS

| Framework | Model | Year | Directed | Signed |
|---|---|---|---|---|
| Information Theory | PIDC | 2017 | × | × |
|  | LEAP | 2017 | × | × |
|  | SCRIBE | 2020 | × | × |
| Machine Learning | GENIE3 | 2010 | × | × |
|  | SCODE | 2017 | ✓ | × |
|  | GRNBoost2 | 2019 | × | × |
|  | PoLoBag | 2020 | ✓ | ✓ |
|  | scSGL | 2022 | × | ✓ |
| Deep Learning | GNE | 2019 | × | × |
|  | CNNC | 2019 | ✓ | × |
|  | DGRNS | 2022 | ✓ | × |
|  | DeepFGRN | 2024 | ✓ | ✓ |
| Graph Deep Learning | GRGNN | 2020 | × | × |
|  | GENELink | 2022 | ✓ | × |
|  | DeepRIG | 2023 | × | × |
|  | GNNLink | 2023 | ✓ | × |
|  | GATCL | 2024 | ✓ | × |
|  | DGCGRN | 2024 | ✓ | × |
|  | CVGAE | 2024 | ✓ | × |
|  | GCLink | 2025 | ✓ | × |
|  | scRegNet | 2025 | ✓ | × |

According to the aforementioned literature (TABLE I), the mainstream approach to GRN inference within the deep learning framework involves integrating single-cell gene expression profiles with established topologies. Among these methods, only PoLoBag [11] and DeepFGRN [22] consider both the direction and sign of gene regulation, while most existing studies focus on only one aspect. However, PoLoBag [11] is based on traditional machine learning and requires manual feature engineering. DeepFGRN [22] uses a GAN network, which does not consider sign and direction based on the topological structure of GRN during feature extraction and therefore does not study the GRN inference problem essentially. As neither PoLoBag [11] nor DeepFGRN [22] is grounded in the fundamental graph-theoretic principle that GRN is inherently a signed directed graph, there exists a inconsistency between the research logic and the biological nature of GRN. This results in gaps in theoretical self-consistency, methodological suitability and the object of research. Consequently, these limitations hinder the development of a coherent study path from "essential cognition" to "accurate inference" of GRN.

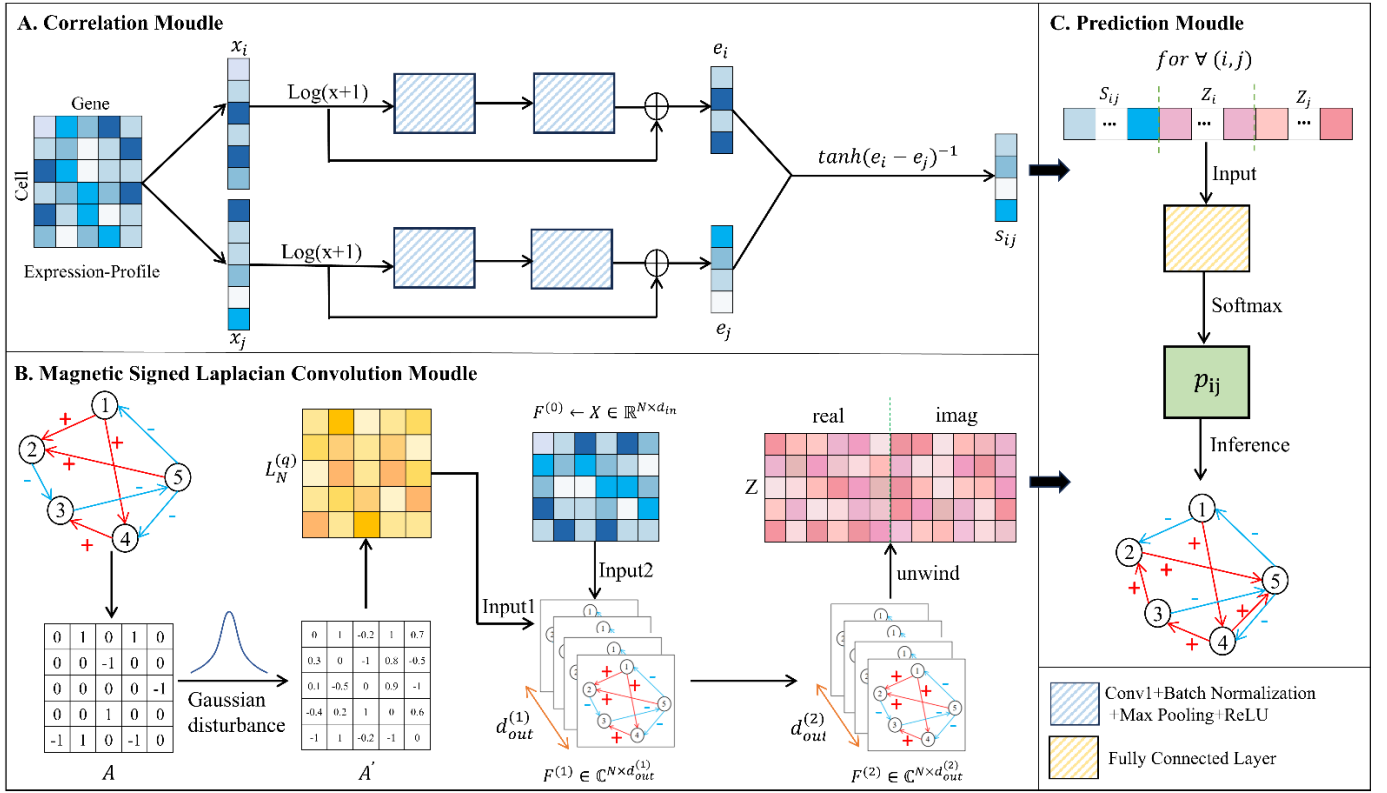

Fig. 1. Overview of the MSGRNLink framework.

To address this gap, we propose a novel approach that explicitly models GRN as a signed directed graph. We introduce the magnetic signed Laplacian convolution, specifically designed for such graphs, enabling a more principled, efficient, and biologically aligned framework for GRN inference.

The primary contributions of this paper are as follows:
1) We develop a framework called MSGRNLink to represent GRN inference as a link prediction task, outperforming current benchmark models on each dataset.
2) Spectral magnetic signed Laplacian convolution is first introduced to GRN inference by modeling GRN as a signed directed graph. This convolution is more elegant and comprehensible, and it has a solid mathematical foundation.
3) We have unified the research logic and the biological nature of GRN. In other words, we directly apply convolution to the topology of the GRN to extract graph features, and the effectiveness of model has been validated in a real medical scenario.

II. METHODS

Since GRN can be viewed as a graph, in this paper we model GRN as a signed directed graph (see leftmost side of Module B in Fig. 1 for a schematic of signed directed graphs) and use deep learning techniques for signed directed graph to infer GRN. We begin by reviewing some fundamental concepts on graph theory and magnetic signed Laplacian matrix.

*A. Definition of directed signed graph*

Signed Directed Graphs (SDGs) are denoted by $G = \{V, \varepsilon^+, \varepsilon^-\}$, where $V = \{v_1, \dots, v_N\}$ is a set containing $N$ nodes, while $\varepsilon^+ \in V \times V$ and $\varepsilon^- \in V \times V$ denote the set of positive and negative edges respectively, and $\varepsilon^+ \cap \varepsilon^- = \emptyset$. These positive and negative edges can also be represented by the signed adjacency matrix $A$, where $A_{i,j} = 1$ if and only if a positive edge exists for $v_i$ and $v_j$, $A_{i,j} = -1$ means that a negative edge exists for $v_i$ and $v_j$, $A_{i,j} = 0$ if there is no edge exists for $v_i$ and $v_j$.

*B. Magnetic Signed Laplacian*

There are several difficulties in creating Laplacian matrix in signed directed graphs: Firstly, the presence of negative edges in signed graphs results in certain elements of the degree matrix being less than or equal to zero, complicating the standardization of the Laplacian matrix and undermining their positive semi-definiteness. Secondly, in directed graphs, conventional Laplacian matrix fails to fulfill the properties of real symmetric matrices, resulting in theoretical adaptability issues during spectral decomposition. The generalized spectral decomposition theorem states that normal matrix in the complex domain can undergo spectral decomposition, with eigenvalues guaranteed to be real if the matrix is Hermitian matrix. Additionally, to ensure the stability of the numerical computation process, we constrain the eigenvalues to the interval [0,2].

He et al [26] develop a convolutional approach specifically for signed directed graphs to tackle the aforementioned problems. The adjacency matrix is redefined, an absolute value degree matrix is utilized, and a phase matrix is established to capture the directional information within the graph as follows:

$$\tilde{A}_{i,j} = \frac{1}{2}(A_{i,j} + A_{j,i}), \quad (1)$$

$$\tilde{D}_{i,i} = \frac{1}{2}\sum_{j=1}^{n}(|A_{i,j}| + |A_{j,i}|), \quad (2)$$

$$\Theta_{i,j}^{(q)} = 2\pi q(A_{i,j} - A_{j,i}), \quad (3)$$

where $q \in \mathbb{R}$ is called the charge parameter.

Utilizing the aforementioned adjacency matrix $\tilde{A}_{i,j}$ and phase matrix $\Theta^{(q)}$, a complex Hermitian matrix $H^{(q)}$ and a normalized magnetic signed Laplace matrix $L_N^{(q)}$ are formulated as follows:

$$H^{(q)} = \tilde{A} \odot exp(i\Theta^{(q)}), \quad (4)$$

$$L_N^{(q)} = I - (\tilde{D}^{-\frac{1}{2}} \tilde{A} \tilde{D}^{-\frac{1}{2}}) \odot exp(i\Theta^{(q)}), \quad (5)$$

where $\odot$ denoting elementwise multiplication and $i$ denoting the imaginary unit. In summary, the core idea of the Hermitian matrix $H^{(q)}$ is that the real part conveys sign information, while the imaginary part conveys direction information.

We employ the Chebyshev filter for modeling, with the filtering operator $Y$ presented as follows:

$$Y = \sum_{k=0}^{K}\theta_k T_k(\tilde{L}_N^{(q)}) = \sum_{k=0}^{K}\theta_k T_k(\frac{2L_N^{(q)}}{\lambda_{max}} - I). \quad (6)$$

III. FRAMEWORK

To utilize signed directed networks to depict gene regulation and infer GRNs, this paper proposes a GRN inference algorithm (MSGRNLink) based on magnetic signed Laplacian convolution. The framework of MSGRNLink is shown in Fig. 1, which consists of three modules: (A) correlation module used to extract similarity features of gene pairs; (B) magnetic signed Laplacian convolution module based on the Chebyshev filter for feature extraction; (C) directed and signed GRN prediction module based on MLP.

A. Correlation Module Based On CNN

The goal of module A is to provide feature representation for the computation of gene pairwise similarity by mapping expression profile of each gene across all samples into a low-dimensional real vector space.

Assuming that there are $N$ genes and $C$ cells in total, the gene expression profile is first normalized by log-transforming it. To obtain the output $A$, input $X \in \mathbb{R}^{C \times N}$ into the first residual block, shown by (7). The output $B$ is then obtained by inputting $A$ into the second residual block, shown by (8). Each residual block consists of a one-dimensional convolutional layer (Conv1), a batch normalization layer (*BN*), a maximum pooling layer (*MP*), and the *RELU* activation function.

$$A = MP(BN(RELU(X * W_1 + b_1))), \quad (7)$$

$$B = MP(BN(RELU(A * W_2 + b_2))), \quad (8)$$

where $*$ is the convolution operation, $W_i$ and $b_i$ are the weight matrix and bias vector of the $i$-th residual block respectively, $i = 1, 2$.

To avoid the problem of gradient vanishing, we introduce skip connections in one-dimensional CNN

$$Shortcut = [MP(MP(BN(X * W_{s_1} + b_{s_1})))] * W_{s_2} + b_{s_2}, \quad (9)$$

where $W_{s_i}$ and $b_{s_i}$ are the weight matrix and bias vector of the $i$-th convolution in the residual connection respectively, $i = 1, 2$.

The corresponding elements of $B$ and *Shortcut* are added together, which is then activated by *RELU* and subjected to Global Average Pooling (*GAP*). Finally, input it into a fully connected layer to obtain the final gene embedding representation $E \in \mathbb{R}^{d_1 \times N}$

$$E = [GAP(ReLU(B + Shortcut))] * W_3 + b_3, \quad (10)$$

where $W_3$ and $b_3$ are the weight matrix and bias vector respectively, and $d_1$ is the final dimension of the node features.

After obtaining the embeddings $e_1, e_2, \ldots, e_N$ of $N$ genes, we need to calculate the similarity of each gene pair. In this paper, we calculate the difference and take the inverse. Then use the $tanh$ function to map it to (-1,1) measuring the similarity of the gene pair $(i, j)$, as shown in (11). The vector $S_{ij}$ preserves both the feature components of the two genes themselves in each dimension and implicitly reflects their similarity in each dimension.

$$S_{ij} = tanh(\frac{1}{e_i - e_j}) \in \mathbb{R}^{d_1 \times 1}. \quad (11)$$

B. Feature Extraction Module Based On Magnetic Signed Laplacian Convolution

Since scRNA-seq data generally approximates a Gaussian distribution post-normalization [27], the dropout resulting from the experimental detection threshold can similarly be regarded as a Gaussian-type measurement error [28]. This paper introduces a minor Gaussian perturbation to the initial adjacency matrix to simulate the continuous fluctuations in gene expression, thereby diminishing the probability of "false zeros" and effectively recovering weak gene expression signals.

Input the gene expression profiles matrix $F^{(0)} \in \mathbb{R}^{N \times d_{in}}$, the complex vectors $F_i^{(L)} \in \mathbb{R}^{1 \times d_{out}^{(L)}}$ and $F_j^{(L)} \in \mathbb{R}^{1 \times d_{out}^{(L)}}$ are derived from the last convolutional layer for the nodes $i$ and $j$. We unwind the $F_i^{(L)}$ and $F_j^{(L)}$ into real-valued vectors $Z_i \in \mathbb{R}^{1 \times 2d_{out}^{(L)}}$ and $Z_j \in \mathbb{R}^{1 \times 2d_{out}^{(L)}}$ respectively.

C. GRN Prediction Module

In the link prediction task (Module C), for any node pair $(i, j)$, to simultaneously consider the similarity between genes and the network topology, we concatenate $S_{ij}$, $Z_i$, and $Z_j$ by rows. The unnormalized scores originate from the fully connected layer, and the edge probability $p_{i,j}$ is predicted using Softmax

$$p_{i,j} = Softmax(concat(S_{ij}, Z_i, Z_j)W), \quad (12)$$

where $W$ represents the weight matrix.

For the GRN with $N$ genes, the total number of possible edges is denoted as $T = N^2$. Suppose $P$ is the probability distribution of the model output, where $p_t$ is the probability vector of the $t$-th sample and $y_t$ signifies the true label. We utilize the Adam optimizer and the negative log-likelihood loss (NLLLoss), as delineated in (13).

$$NLLLoss(P, y) = -\frac{1}{T}\sum_{t=1}^{T} p_t y_t. \quad (13)$$

## IV. EXPERIMENTS

### A. Configuration of Model Parameters in MSGRNLink

We conceptualize GRN inference as a link prediction task within signed directed graphs. For nodes $i$ and $j$, there are five classification cases, respectively $(i,j) \in \mathcal{E}^+$, $(i,j) \in \mathcal{E}^-$, $(j,i) \in \mathcal{E}^+$, $(j,i) \in \mathcal{E}^-$ and $(i,j) \notin \mathcal{E}, (j,i) \notin \mathcal{E}$, as seen in Fig. 2.

The parameter configurations of module A are presented in TABLE II, and we experimentally determine the optimal combinations $K$, $q$, and $L$ in module B. The $K$ is the number of terms in the Chebyshev polynomial expansion. The charge parameter $q$ in the phase matrix $\Theta^{(q)}$ influences the direction of the phase alteration. The $L$ represents the quantity of convolutional layers, which affects the feature extraction capability of the model. We conduct experiments on dataset E to determine the optimal experimental configuration. Fig. 3 illustrates the experimental procedures, whereas Fig. 4 presents the experimental results.

Initially, we fix $K = 1$ and $L = 2$, modify the value of $q$ (set $q_0 = \frac{1}{2max_{i,j}(A_{i,j} - A_{j,i})}$). The vast majority of metrics are all optimal at $q = q_0 \times \frac{1}{5}$, so we set $q = q_{optimal} = q_0 \times \frac{1}{5}$ in the subsequent experiments of this work.

Subsequently, we fix $q = q_0 \times \frac{1}{5}$ and $L = 2$, and alter the value of $K$. When $K = 1$, all the evaluation metrics exhibit their maximum values. Consequently, in the subsequent experiments detailed in this paper, we set $K = 1$.

Ultimately, we fix $q = q_0 \times \frac{1}{5}$ and $K = 1$, modify the value of $L$. When $L = 2$, all indices reach the maximum value. Therefore, in the following experiments presented in this paper, we set $L = 2$.

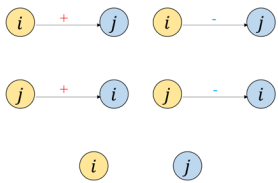

Fig. 2. Five prediction results.

TABLE II
THE PARAMETER SETTING OF MODULE A

| Procedure | Number | Operation | Parameter setting |
|---|---|---|---|
| Main branch | 1 | Conv1d | in=1, out=16, kernel=9, padding=4 |
| | 2 | BN | channel=16 |
| | 3 | MP | kernel=2, stride=2, padding=1 |
| | 4 | ReLU | / |
| | 5 | Conv1d | in=16, out=16, kernel=5, padding=2 |
| | 6 | BN | channel=16 |
| | 7 | MP | kernel=2, stride=2, padding=1 |
| | 8 | ReLU | / |
| Skip connection | 1 | Conv1d | in=1, out=16, kernel=1, padding=0 |
| | 2 | BN | channel=16 |
| | 3 | MP | kernel=2, stride=2, padding=1 |
| | 4 | MP | kernel=2, stride=2, padding=1 |
| | 5 | Conv1d | in=16, out=16, kernel=1, padding=0 |
| Add | 1 | GAP | out=1 |
| | 2 | Squeeze | / |
| | 3 | Linear | in=16, out=$d_1$ |

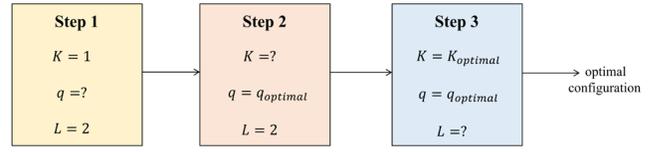

Fig. 3. Determine the optimal configuration of parameters in module B.

### B. Benchmark datasets

In order to test the GRN inference ability of MSGRNLink, we conduct experiments on both simulated datasets and real datasets. In accordance with [11], five simulated datasets A-E were utilized in this paper. We employed nine real directed signed network datasets. Dataset I was a subnetwork constructed from human GRN in TRRUST v2 database [29], with gene expression data sourced from the GEO database with accession number GDS3795, from bone marrow CD34+ cells of myelodysplastic syndrome patients and healthy controls [30]. We downloaded the gene expression data for E.coli and human diseases from the Gene Expression Omnibus (GEO) [31] and Gene Expression Nebulas (GEN) databases [32], respectively. Gene pairs with specified directionality and regulatory types were compiled from RegulonDB [33], TRRUST V2 [34], and RegNet-work [35]. The E.coli expression data encompass numerous conditions, including cold stress, heat stress, lactose stress, and oxidative stress. The human disease dataset consists of four types: breast cancer, Coronavirus disease 2019 (COVID-19), liver cancer, and lung cancer. The details of the datasets are presented in TABLE III.

### C. Performance metrics

We select the Area Under the Receiver Operating Curve ($AUROC$) as the definitive evaluation metric, where $AUROC$ represents the area under the ROC curve. The definition of $AUROC$ is given as follows:

$$AUROC = \frac{TP + TN}{TP + FP + TN + FN}, \quad (14)$$

where $TP$ denotes the number of correctly identified links, $TN$ denotes the number of correctly identified false links, $FP$

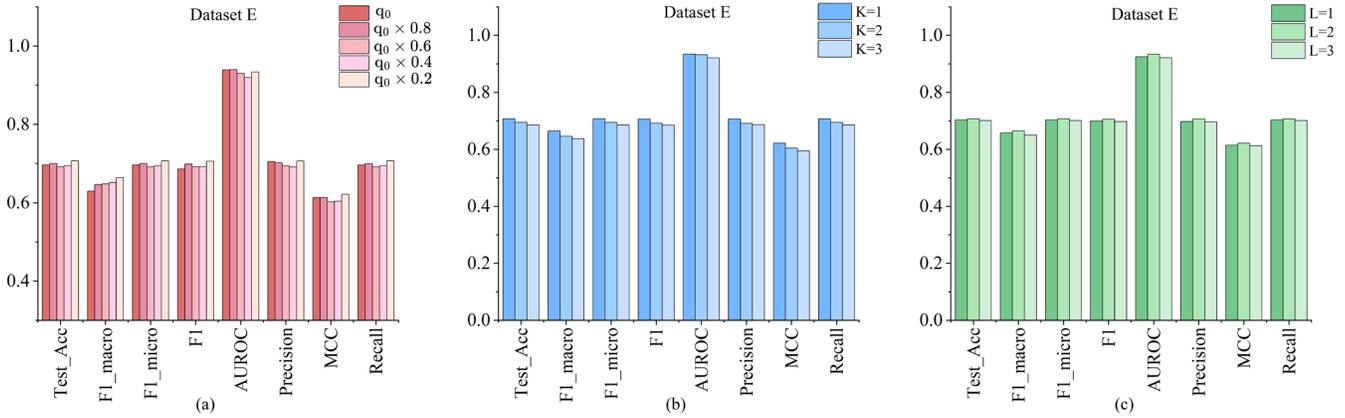

Fig. 4. Experimental results for changing parameters in module B. (a) Fixed $K$ and $L$, change $q$. (b) Fixed $q$ and $L$, change $K$. (c) Fixed $K$ and $q$, change $L$.

TABLE III
BENCHMARK DATASETS

| Dataset | Type | Organism | Directed | Positive Edges | Negative Edges | G_num | C_num | Known_Edge | TF | Target | Density |
|---|---|---|---|---|---|---|---|---|---|---|---|
| A | Simulated | Yeast | Yes | 238 | 237 | 200 | 200 | 475 | 17 | 197 | 0.0119 |
| B | Simulated | Yeast | Yes | 532 | 539 | 400 | 400 | 1071 | 49 | 385 | 0.0067 |
| C | Simulated | Yeast | Yes | 872 | 950 | 500 | 500 | 1822 | 96 | 478 | 0.0073 |
| D | Simulated | E.coil | Yes | 812 | 568 | 500 | 500 | 1380 | 64 | 482 | 0.0055 |
| E | Simulated | E.coil | Yes | 886 | 637 | 650 | 650 | 1523 | 66 | 632 | 0.0036 |
| I | Real | Human | Yes | 1283 | 762 | 408 | 200 | 2045 | 253 | 338 | 0.0123 |
| Ecoli_cold | Real | E.coil | Yes | 2308 | 2261 | 2205 | 24 | 4569 | 174 | 2007 | 0.0009 |
| Ecoli_heat | Real | E.coil | Yes | 2308 | 2261 | 2205 | 24 | 4569 | 174 | 2007 | 0.0009 |
| Ecoli_lactose | Real | E.coil | Yes | 2308 | 2261 | 2205 | 12 | 4569 | 174 | 2007 | 0.0009 |
| Ecoli_oxidative | Real | E.coil | Yes | 2308 | 2261 | 2205 | 33 | 4569 | 174 | 2007 | 0.0009 |
| human_breast | Real | Human | Yes | 6919 | 1988 | 2478 | 24 | 8907 | 756 | 2074 | 0.0015 |
| human_COVID | Real | Human | Yes | 6919 | 1988 | 2478 | 42 | 8907 | 756 | 2074 | 0.0015 |
| human_liver | Real | Human | Yes | 6919 | 1988 | 2478 | 10 | 8907 | 756 | 2074 | 0.0015 |
| human_lung | Real | Human | Yes | 6919 | 1988 | 2478 | 130 | 8907 | 756 | 2074 | 0.0015 |

denotes the number of incorrectly identified links, and *FN* denotes the number of incorrectly identified false links.

*D. Parameter analysis*

The parameters have a strong influence on the performance of the model, so we carry out tuning experiments on these parameters on dataset E, and the experimental results are shown in Fig. 5.

First, we analyze the output dimension $d_1$ of module A. At a dimension of 64, many key evaluation metrics reach their optimal value, including Test_Acc, et al. Therefore, $d_1 = 64$ is the optimal choice.

Second, we analyze the hidden dimension in module B. Analysis indicates that dimension 48 attains the highest values for most of critical metrics and is preferred.

Next, we analyze the impact of the dropout ratio on the model performance. As the dropout ratio is gradually raised from 0% to 60%, most evaluation metrics show a consistently increasing trend. When the dropout ratio continues to increase to 70%, all indicators decrease. As a result, a dropout percentage of 60% is employed in this paper.

Ultimately, we study the performance across various training set ratios. To balance training efficacy and testing stability, we select an 80% training set ratio in this paper.

Furthermore, we investigated the generalization capacity of the model for gene and cell numbers across various scales, with the experimental results shown in Fig. 6(a-b). The results show that our model is more appropriate for medium-sized networks. We also plot the loss and accuracy curves on the training set to assess the stability of model, taking dataset A as an example, with the experimental results shown in Fig. 6(c). The loss curve rapidly decreases and stabilizes at a low value, indicating that the model effectively optimizes parameters and converges. The accuracy curve increases quickly and maintains a high level, demonstrating strong predictive capability and stability throughout the training process.

*E. Ablation Study*

This section conducts six ablation experiments to assess the efficacy of each module in the MSGRNLink. We implemented 5-fold cross-validation (FCV) methods 10 times, Fig. 7 shows the experimental outcomes for dataset A. The experimental settings are outlined as follows:
• E.1: The model proposed in this paper (MSGRNLink).
• E.2: Remove gaussian perturbation from E.1.
• E.3.1: Substitute module A with MLP based on E.1.
• E.3.2: Substitute module A with the Spearman correlation coefficient based on E.1.
• E.3.3: Substitute module A with the dot product based on E.1.
• E.4: Based on E.1, replace module B with a simple Graph Convolution Network (GCN). Given that $A$ and $D$ are the

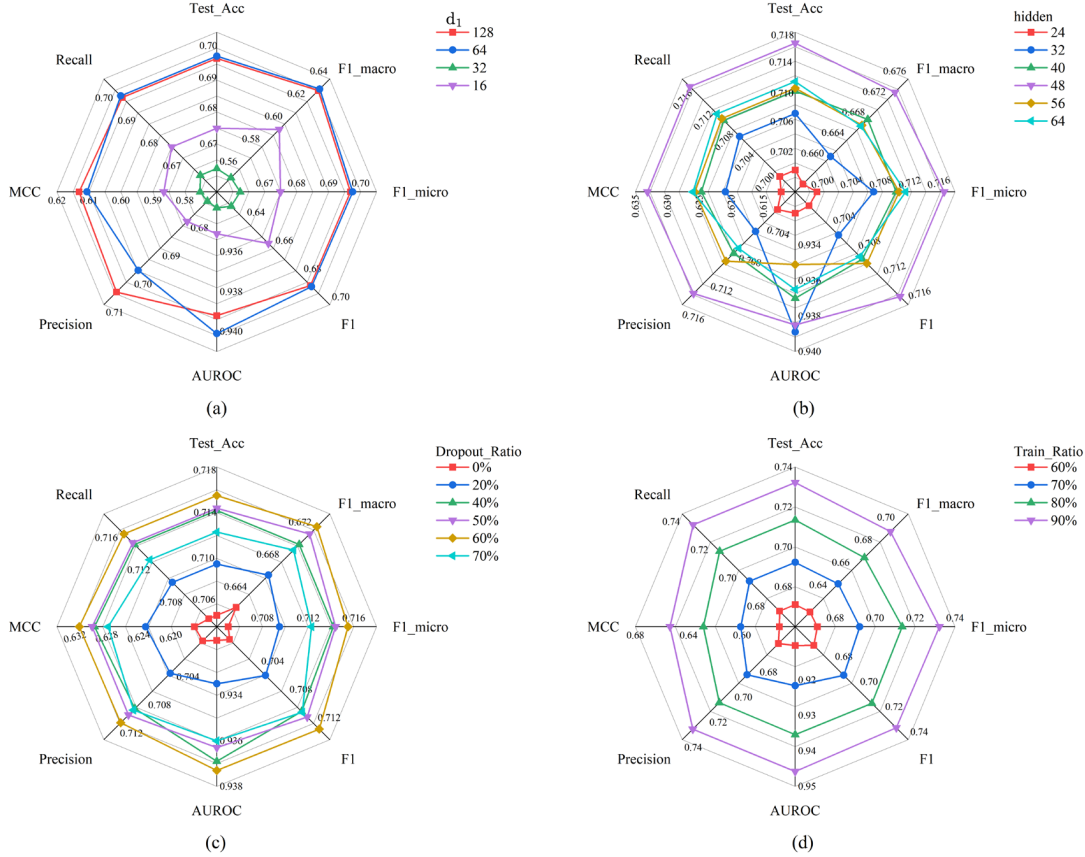

Fig. 5. Experimental results of parametric sensitivity analyses. (a) $d_1$. (b) hidden. (c) Dropout_Ratio. (d) Train_Ratio.

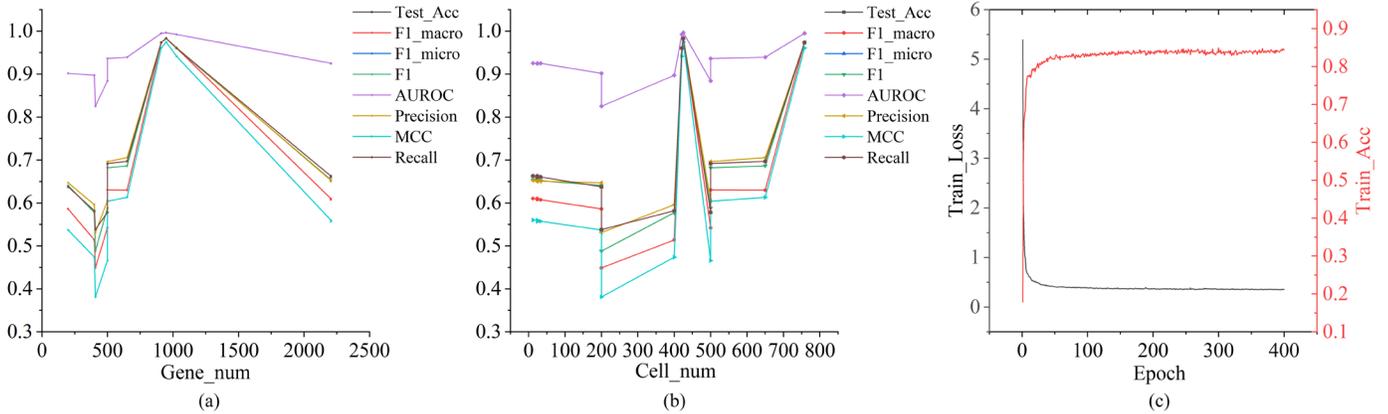

Fig. 6. (a) Model performance at different gene scales. (b) Model performance at different cell scales. (c) Training loss and accuracy curves over epochs for dataset A.

adjacency matrix and degree matrix of graph $G$ respectively, the embeddings of the $\ell$th layer are characterized by $H^{(\ell)} = ReLU(D^{-0.5}AD^{-0.5}H^{(\ell-1)}W^{(\ell-1)} + b^{(\ell-1)})$, where W and b are the weight matrix and bias vector respectively.

Upon individual comparison of E.1 and E.2, it is observed that the removal of the Gaussian perturbation results in a slight growth of all metrics. Considering the significance of Gaussian perturbation, we retain it in our experiment.

Upon individual comparison of E.1 and E.3.1, E.1 and E.3.2, E.1 and E.3.3, it is evident that most metrics exhibit a downward trend, accompanied by an increase in dispersion.

In addition, among all ablation experiments, E.4 has the worst performance and the largest data fluctuation (e.g., Recall has a wide distribution of outliers). These results suggest that the traditional Laplacian matrix is ill-suited for the GRN inference task, thereby underscoring the effectiveness of the magnetic signed Laplacian matrix.

In conclusion, both the magnetic signed Laplacian convolution module and the similarity feature extraction module proposed in this paper contribute positively to the model performance, demonstrating that the complete model architecture is more suitable for the GRN inference task.

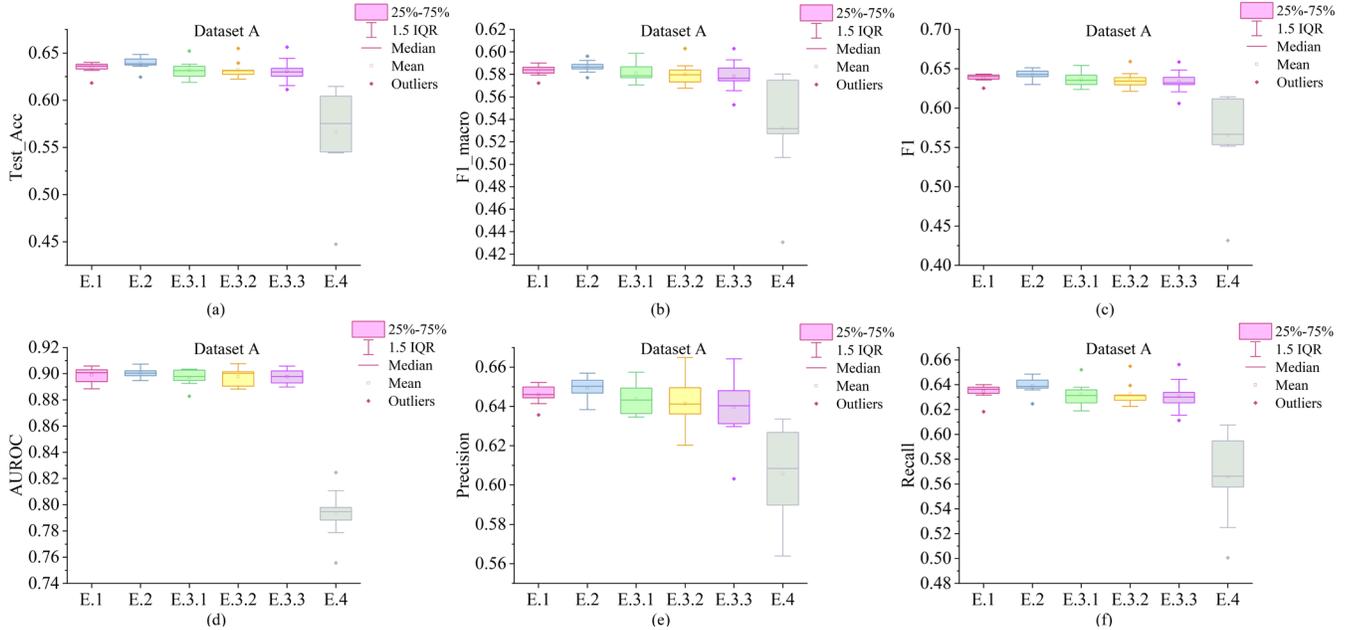

Fig. 7. The results of ablation experiments. (a) Test_Acc. (b) F1_macro. (c) F1. (d) AUROC. (e) Precision. (f) Recall.

*F. Comparing with other methods*

To assess the performance of MSGRNLink in GRN inference, we select several state-of-the-art GRN inference algorithms in deep learning for comparison, including DeepRIG [19], DeepFGRN [22], GNNLink [20] and GENELink [17], with the evaluation metric been the AUROC. We conduct five-fold cross-validation (FCV) 10 times on both simulated and real datasets, then average the results, as shown in Fig. 8.

To ensure a fair comparison in the experiments, we adjust the number of negative samples in the code of DeepFGRN [22] (default 6000) to match the quantity of known edges in the ground truth of dataset. Excessive negative samples may dominate the loss function and gradient updating, triggering prediction bias. Therefore, maintaining a balanced ratio of positive to negative samples (1:1) is essential for experimental design to ensure the reliability of the GRN.

Fig. 8(a) illustrates that MSGRNLink earns the highest AUROC values across all datasets, markedly surpassing other models. The 3D bar chart (Fig. 8(b-d)) further visualizes the superior predictive performance of MSGRNLink. The 3D line graph (Fig. 9) further demonstrates that MSGRNLink has the least fluctuations in AUROC folds, highlighting the robustness across data scenarios.

In addition to the prediction performance, we also computed the running time of different models across all datasets to evaluate their computational complexity, the results are listed in TABLE IV. Although MSGRNLink involves intricate spectral-domain convolution and complex operations, it does not incur the highest computational cost. On the contrary, it outperforms most competing models, highlighting the efficiency and cost-effectiveness.

## V. CASE STUDY

GRN is a complex network wherein diseases may arise from mutations and dysregulation of genes, regulatory elements, and pathways, thereby disturbing the physiological balance of the organism [36]. Accordingly, we conduct a case study on the BLCA dataset (sourced from [21]) to demonstrate the effectiveness and biological significance of the model MSGRNLink.

Fig. 10 (a) shows the known ground truth of the BLCA dataset, while Fig. 11(a) presents the prediction results of DGCGRN [21], Fig. 11(b) displays the prediction results of DeepRIG [19] (the threshold is chosen to be 0.6 to determine the presence of edges), Fig. 11(c) depicts the prediction results of DeepFGRN [22], and Fig. 11(d) showcases the prediction results of MSGRNLink. (Solid lines represent known edges, while dashed lines represent newly predicted edges.) Neither DGCGRN [21] nor DeepRIG [19] is capable of sign prediction, and notably, DeepRIG [19] mistakenly infers the regulatory direction between MAP2K1 and MAPK1 compared with Fig. 10(a). Although DeepFGRN [22] can forecast the direction and sign, it ignores the contributions of three genes (RPS6KA5, MDM2, and TP53). Conversely, MSGRNLink predicts more effectively and comprehensively.

TABLE IV
RUNNING TIME OF DIFFERENT MODELS ON ALL DATASETS

| Datasets | MSGRNLink | DeepRIG | DeepFGRN | GNNLink | GENELink |
|---|---|---|---|---|---|
| A | 40s | 9s | 3min53s | 16s | 10min06s |
| B | 1min | 19s | 4min33s | 15s | 14min14s |
| C | 1min41s | 27s | 6min3s | 20s | 13min53s |
| D | 1min36s | 28s | 5min28s | 15s | 13min24s |
| E | 1min40s | 33s | 6min13s | 16s | 15min17s |
| I | 1min39s | 22s | 5min50s | 13s | 7min56s |
| Ecoli_cold | 6min3s | 3min9s | 12min56s | 35s | 12min15s |
| Ecoli_heat | 7min59s | 2min37s | 11min5s | 43s | 12min40s |
| Ecoli_lactose | 8min43s | 2min38s | 10min11s | 47s | 12min39s |
| Ecoli_oxidative | 8min34s | 3min18s | 16min36s | 48s | 12min7s |
| human_breast | 11min54s | 4min4s | 22min14s | 1min20s | 13min10s |
| human_COVID | 9min44s | 3min57s | 17min49s | 58s | 13min7s |
| human_liver | 12min10s | 3min24s | 21min23s | 1min25s | 13min58s |
| human_lung | 12min56s | 4min11s | 27min53s | 1min27s | 13min46s |

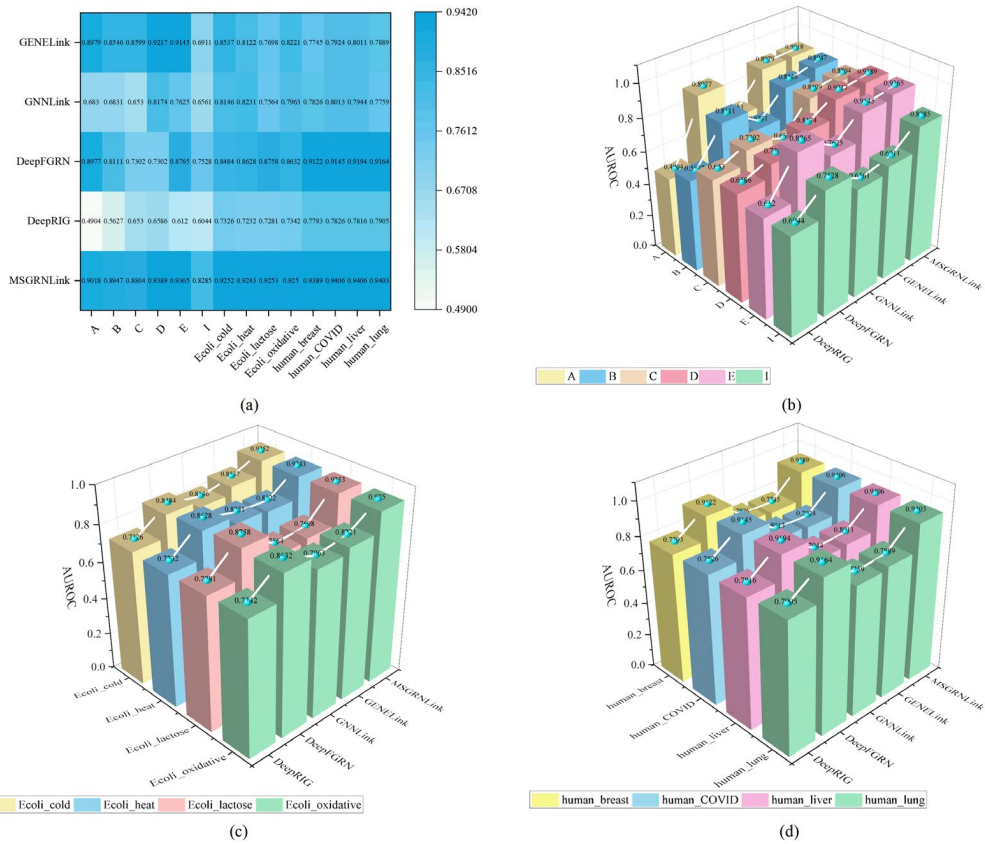

Fig. 8. Comparison AUROC of MSGRNLink with existing state-of-the-art methods. (a) Heat map with labels. (b) 3D bar chart on A-E and I datasets. (c) 3D bar chart on Ecoil datasets. (d) 3D bar chart on human datasets.

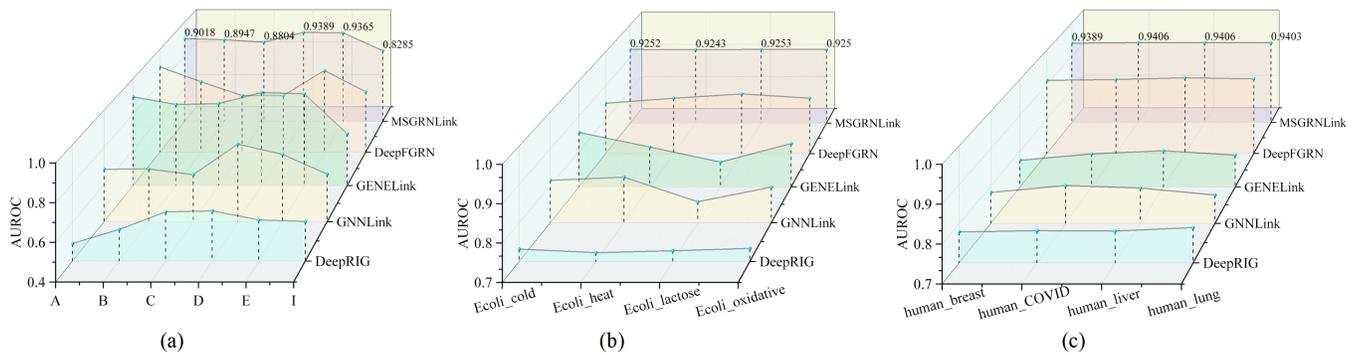

Fig. 9. 3D Line Chart. (a) A-E and I datasets. (b) Ecoil datasets. (c) Human datasets.

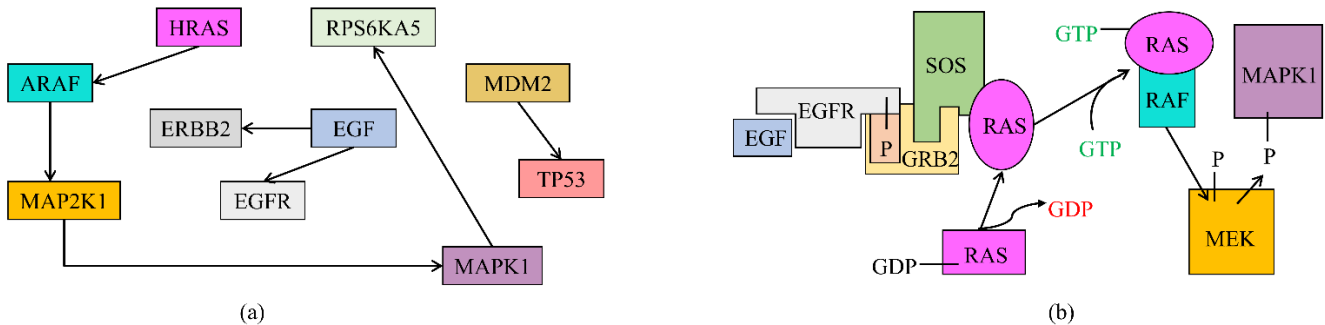

Fig. 10. (a) Ground Truth. (b) MAPK signal pathway.

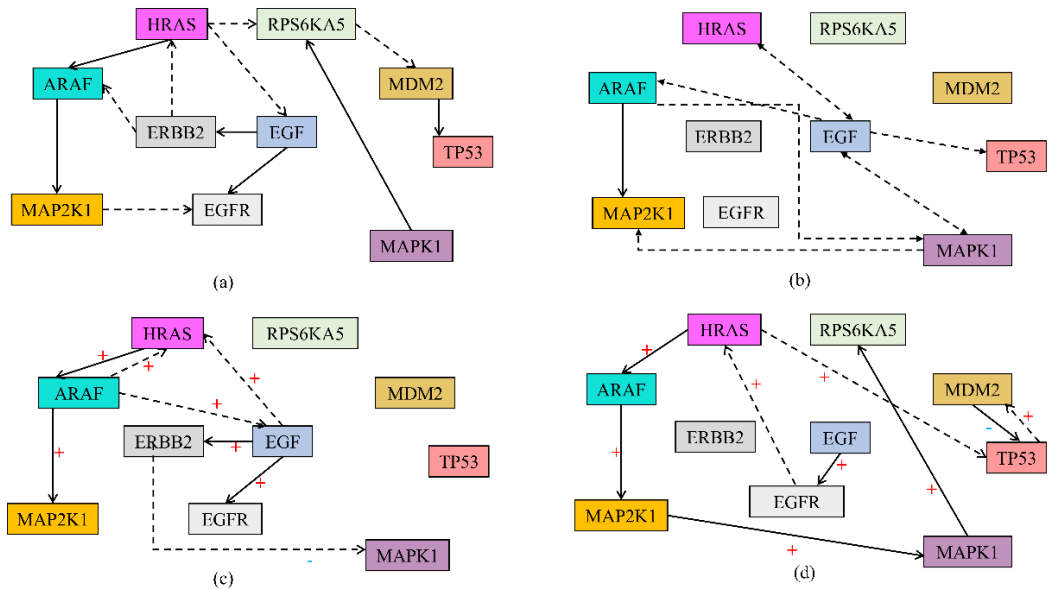

Fig. 11. (a) GRN inferred by DGCGRN. (b) GRN inferred by DeepRIG. (c) GRN inferred by DeepFGRN. (d) GRN inferred by MSGRNLink.

We identified the MAPK signal pathway, shown in Fig. 10(b), which has been confirmed by the GeneCards database and WIKIPEDIA platform. Epidermal growth factor (EGF) initially attaches to the receptor (EGFR) on the cell membrane, resulting in its activation. Upon EGFR activation, the inactive GDP of the RAS protein is exchanged for active GTP. The binding of RAS to GTP activates the RAF kinase, which then phosphorylates and activates MAP2K1 (often referred to as MEK), leading to the phosphorylation and activation of MAPK1 (also known as ERK2) by MAP2K1. Furthermore, [37] suggests that EGF-induced activation of EGFR initiates multiple downstream signaling cascades, among which the MAPK (RAS-RAF-MEK-ERK) pathway plays a dominant role. As noted in [37], this EGF-induced signaling pathway serves as a novel regulatory mechanism that orchestrates the synchronized expression of a specific subset of functionally related genes to facilitate effective cell proliferation. Aberrant activation of the EGF-EGFR-ERK pathway drives the uncontrolled proliferation of cancer cells.

Moreover, we identified and accurately predicted a known relationship between MAPK1 and RPS6KA5 with a positive sign, and this finding is supported by "UCSC Genome Browser" platform. Specifically, MAPK1 can directly phosphorylate multiple key sites of RPS6KA5 (also known as MSK1), thereby activating its N-terminal kinase structural domain and allowing RPS6KA5 to acquire catalytic function. The activated RPS6KA5 can in turn phosphorylate a variety of downstream substrates, thereby regulating the expression of pro-proliferative/survival genes. The ERK1/2 pathway is frequently over-activated in various tumor types, resulting in MSK1 continuously phosphorylating downstream targets. This process promotes tumor cell proliferation, survival, and epithelial–mesenchymal transition (EMT), thereby enhancing metastatic potential. [38] has demonstrated that the activation of MSK1 can promote EMT, thereby significantly enhancing the metastatic potential of colon cancer cells.

Strikingly, our model also predicted a positive feedback loop between Tp53 and HRAS and a negative feedback loop between Tp53 and MDM2. [39] experimentally demonstrated that wild-type p53, either over-expressed or activated in response to cellular stress, directly binds and activates HRAS, thereby upregulating HRAS transcription and mRNA levels. In addition, the activated Ras signal, in turn, helps to maintain the stability of p53, forming a "positive feedback loop". [40] investigated the negative feedback relationship between p53 and MDM2. After DNA damage, stabilized/activated p53 binds to the P2 promoter of MDM2 and promotes its transcription [41], and MDM2 represses p53 in turn.

In conclusion, the comparison of Fig. 11(a-d) demonstrates that we not only predicted the majority of established regulations but also clarified the sign (activation/inhibition), all of which are validated by existing literature or databases, thereby underscoring the biological worth of MSGRNLink.

## VI. CONCLUSION

Inferring GRNs from gene expression data and network topologies has been a key challenge in systems biology, aiding in the identification of disease biomarkers and the comprehension of gene regulatory mechanisms. Nonetheless, previous studies have overlooked the graph-theoretic foundation of treating GRNs as signed directed graphs, resulting in research logic that deviates from the intrinsic nature of GRNs. This misalignment hampers the accuracy and interpretability of GRN inference.

Therefore, in this paper, we propose MSGRNLink, a gene regulatory network inference model that extracts node features through magnetic signed Laplacian convolution tailored for signed directed graphs and employs a correlation module for identifying potential gene similarities, framing GRN inference as a link prediction task. Experimental results on five simulated datasets and nine real datasets demonstrate that MSGRNLink outperforms all benchmark models and is computationally efficient. Additionally, we conduct a parameter sensitivity analysis to illustrate the robustness of the model. Ablation experiments show the significance of the magnetic signed Laplacian convolution module in the GRN inference task, as it captures both direction and sign information, which is more relevant to real situations. The case study on the real bladder urothelial carcinoma (BLCA) also demonstrates that MSGRNLink can not only correctly infer known regulatory relationships but also predict the regulatory signs (with positive signs indicating activation and negative signs indicating inhibition). Moreover, all predicted regulatory relationships are supported by factual evidence, confirming their validity. This contributes to the development of therapeutic drugs for related diseases and holds significant clinical relevance.

Of course, MSGRNLink also has certain defects. First, adding Gaussian perturbations to the initial adjacency matrix in module B reduces the probability of "false zeros" appearing, but many "true zeros" are also perturbed, which can be further improved. Second, module B of MSGRNLink does not incorporate biological features during feature extraction, such as ATAC-seq features or RNA-seq features. The two limitations mentioned above will be the focus of our future work.

ACKNOWLEDGMENT

The authors would like to thank all authors of the cited references.

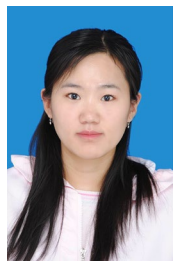

**Wei Xiong** received the PhD degree from Xinjiang University, in 2014. She is currently an associate professor with the School of Mathematics and Systems Science, Xinjiang University. Her major research interests include graph theory and application.

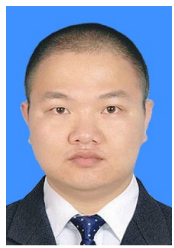

**Yuannong Ye** received the PhD degree from University of Electronic Science and Technology of China in 2015. He is currently an associate professor with the School of Biology and Engineering, Guizhou Medical University. His major research interests include bioinformatics.

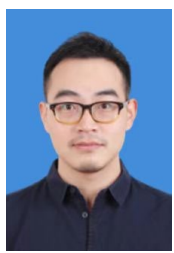

**Bin Zhao** received the PhD degree from Xinjiang University, in 2016. He is currently an associate professor with the School of Biology and Engineering, Guizhou Medical University. His major research interests include deep learning.

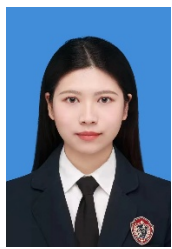

**Rijie Xi** received the BA degree in Information Management and Information Systems from Nanjing University of Posts and Telecommunications, China, in 2023. She is currently working toward the master's degree in Applied Statistics at the School of Mathematics and Systems Science, Xinjiang University. Her research interests include bioinformatics.

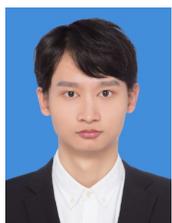

**Weikang Xu** received the BS degree in Applied Statistics from the College of Mathematics and Computer Science, Zhejiang A&F University, China, in 2023. He is currently pursuing the master's degree in Applied Statistics at the School of Mathematics and Systems Science, Xinjiang University. His research interests include deep learning.